\documentclass[journal]{IEEEtran}
\usepackage{cite} 
\usepackage{amsmath} 
\usepackage{amsfonts,amssymb} 
\usepackage{mathrsfs} 
\usepackage{graphicx,epsfig,balance} 
\usepackage[utf8]{inputenc} 
\usepackage{CJK} 
\usepackage{bm} 
\usepackage{algorithm} 
\usepackage{epsfig}
\usepackage{algorithmic}
\usepackage{extarrows} 
\usepackage{color} 
\usepackage{ragged2e} 
\usepackage{makecell}
\usepackage{indentfirst}
\usepackage{amsmath}

\newcommand{\tu}{\textup}

\begin{document}
\title{Deep-Learning-based Millimeter-Wave Massive MIMO for Hybrid Precoding}

\author{ Hongji Huang, \emph{Member}, \emph{IEEE}, Yiwei Song, Jie Yang, \emph{Member}, \emph{IEEE}, \\ Guan Gui, \emph{Senior Member}, \emph{IEEE}, and Fumiyuki Adachi, \emph{Life Fellow}, \emph{IEEE}
\thanks{H. Huang, Y. Song, J. Yang, and G. Gui are with College of Telecommunication and Information Engineering, Nanjing University of Posts and Telecommunications, Nanjing 210003, P. R. China. (e-mail:
\mbox{b14111829@njupt.edu.cn}, {b15080234@njupt.edu.cn}, {guiguan@njupt.edu.cn}, hikmet@njupt.edu.cn)}
\thanks{F. Adachi is with Wireless Signal Processing Research Group, Research Organization of Electrical Communication (ROEC), Tohoku University, Sendai 980-8577, Japan. (e-mail: \mbox{adachi@ecei.tohoku.ac.jp})}
}

\maketitle
\begin{abstract}
Millimeter wave (mmWave) massive multiple-input multiple-output (MIMO) has been regarded to be an emerging solution for the next generation of communications, in which hybrid analog and digital precoding is an important method for reducing the hardware complexity and energy consumption associated with mixed signal components. However, the fundamental limitations of the existing hybrid precoding schemes is that they have high computational complexity and fail to fully exploit the spatial information. To overcome these limitations, this paper proposes a deep-learning-enabled mmWave massive MIMO framework for effective hybrid precoding, in which each selection of the precoders for obtaining the optimized decoder is regarded as a mapping relation in the deep neural network (DNN). Specifically, the hybrid precoder is selected through training based on the DNN for optimizing precoding process of the mmWave massive MIMO. Additionally, we present extensive simulation results to validate the excellent performance of the proposed scheme. The results exhibit that the DNN-based approach is capable of minimizing the bit error ratio (BER) and enhancing spectrum efficiency of the mmWave massive MIMO, which achieves better performance in hybrid precoding compared with conventional schemes while substantially reducing the required computational complexity.

\end{abstract}

\begin{IEEEkeywords}
Millimeter wave (mmWave), massive multiple-input multiple-output (MIMO), deep learning, hybrid precoding.
\end{IEEEkeywords}

%
\IEEEpeerreviewmaketitle

\section{Introduction}

\IEEEPARstart{W}{ireless} data traffic is predicted to improve 1000-fold by the year 2020 and may increase by over 10 000-fold by the year 2030 \cite{A2014}, thus promoting the development of the fifth generation (5G) concept to cope with the explosive data increase. As one of the most highly efficient techniques to meet the 5G requirements, the use of enormous chunks of under-utilized spectrum in the ultra-high-frequency bands, such as the millimeter-wave (mmWave) band, has recently attracted considerable interest in the research community \cite{w2014}. Compare with current wireless systems, one evolutionary progress of mmWave communications is that the ten-fold increase in carrier frequency. In another words, mmWave signals bring an orders-of-magnitude enhancement in free-space pathloss \cite{o2014}.

Inspired by massive multiple-input multiple-output (MIMO), a mmWave massive MIMO system is considered to be a potential technique for enhancing system throughput. To multiplex a large amount of data streams and achieve more accurate beamforming in mmWave massive MIMO, hybrid precoding was proposed \cite{A2014}. In \cite{gao2016}, a successive interference cancellation-based hybrid precoding that can realize excellent performance with low complexity was presented, in which the sum rate optimization problem with non-convex constraints was divided into several sub-rate optimization issues. Then, the authors in \cite{A2015} designed a low-complexity hybrid analog/digital precoding for multiuser mmWave systems by configuring a hybrid precoder. However, these precoding schemes proposed in the previous works have high commotional complexity and require a complicated bit allocation strategy since the previously proposed hybrid anlog/digital precoding schemes are based on singular value decomposition (SVD). Additionally, the newly proposed geometric mean decomposition (GMD)-based scheme \cite{chen2015} can avoid the bit allocation issue, but it still brings great challenges in addressing the non-convex constraint on the analog precoder and in exploiting the structural characteristics of the mmWave massive MIMO systems.

However, in the context of the  mmWave massive MIMO systems, though a quantity of researches have been devoted to enhancing the hybrid precoding performance, there are still a lot of problems remaining and two major challenge are the extraordinarily high computational complexity and poor system performance. In the past few years, many scholars have realized this gap and they have provided different methods for reducing the computational complexity or improving the precoding performance, such as the (GMD)-based scheme \cite{chen2015}, the matrix factorization-based hybrid precoding mean \cite{rusa}, a precoding method based on radio-frequency (RF) and baseband signal processing \cite{rf}, and hybrid spatial processing architecture aided precoding approach \cite{gwang}, et al. Also, for the sake of realizing high spectrum efficiency with low complexity, an alternating minimization scheme for effectively designing hybrid precoder was provided \cite{jstp}. Then, by exploiting low-dimensional beamspace channel state information (CSI) processed by compressive sensing (CS) detectors, paper \cite{ntw} presented a beamspace-SVD based hybrid precoding method for reducing complexity. In general, these works are based on conventional mathematical means such as the SVD and the GMD, which are too weak to exploit the sparsity statistics of the mmWave massive MIMO. Simultaneously, since the traditional methods are inadequate to leverage the structural characteristics of such mmWave systems, traditional low-complexity schemes are realized at the cost of degrading the hybrid precoding of the systems. Therefore, previous works fail to deal with these issues fundamentally and new methods are urgent need to be put forward for enhancing the hybrid precoding performance of the mmWave massive MIMO.

Recently, the emerging solution called deep learning \cite{Hinton2006} is an extraordinarily remarkable technology for handling explosive data and addressing complicated nonlinear problems. It has been proved that deep learning is an excellent tool to deal with complex non-convex problems and high-computation issues, which is dedicated by its super-excellent recognition and representation abilities. Some previous works which incorporate deep learning into communications have been investigated, including beam selection, heterogeneous network, non-orthogonal multiple access (NOMA), massive MIMO, and heterogeneous network \cite{f2017,n2017,b2017,g2018,g2,g3}. Additionally, deep learning has been applied to intelligent traffic control area \cite{ft,zm,hg,ft2}, showing great advancements resulting from the deep-learning-based communication schemes.

Thus, this study investigates a framework which integrates deep learning into hybrid precoding in mmWave MIMO systems. The main contributions of this paper are summarized as follows.
\begin{enumerate}
  \item First, this is the first work to design a framework that incorporates the deep learning technique into hybrid precoding. Specifically, we regard a deep neural network (DNN) as an autoencoder, and this model is regarded as a black box, where activation functions optimize multiple layers of the network and create corresponding mapping relations.
  \item In our work, a hybrid precoding scheme based on deep learning is provided. Here, the DNN is capable of capturing structural information of hybrid precoding scheme through the training stage, contributing to lowering the computational complexity. Additionally, simulation results and comparisons have verified the superiority of the proposed methods.
\end{enumerate}

The rest of this paper is organized as follows. To begin with, we establish a mmWave massive MIMO model, in which many antennas are implemented at the base station (BS). Then, in Section III, to achieve hybrid precoding with good performance, we develop a DNN framework and provide a deep learning-enabled scheme. Simulation results for assessing the performance of the deep learning-based method are provided in Section IV, and conclusions are presented in Section V.

\emph{Notations:} $N_s$ is denoted as independent data streams, $N_t$ and $N_r$ are defined as the transmitted antennas and the received antennas. Also, $N_t^{RF}$ and $N_r^{RF}$ represent RF chains. Furthermore, $N$ is denoted as the number of samples.

\section{System Model}
\label{sec2}

We consider a typical mmWave massive MIMO system, in which one BS with a uniform linear array (ULA) of $N_t$ antennas and user with $N_r$ received antennas are designed. Here, the BS sends $N_s$ independent data streams to the user, and it is assumed to have no information on all the communication links. Additionally, it is assumed that the BS and the user have $N_t^{RF}$ and $N_r^{RF}$ RF chains, respectively, which meet the requirements that $N_s\leq N_t^{RF}\leq N_t$ and $N_t\leq N_r^{RF}\leq N_r$ \cite{o2014}. Furthermore, we introduce the well-known Saleh-Valenzuela (SV) channel model \cite{o2014}, and the channel matrix $\mathbf H\in \mathbb{C}^{N_t\times N_r}$ is written by
\begin{align}
\label{h1}
\mathbf{H} = \sqrt{\frac{N_t N_r}{P}} \left(\alpha_0 \mathbf{a}_t(\theta_{0}^t) \mathbf{a}_r(\theta_0^r) + \sum_{p=1}^P \alpha_p \mathbf{a}_t(\theta_{p}^t) \mathbf{a}_r(\theta_p^r)\right),
\end{align}
Here, $P$ denotes the number of non-line-of-sight (NLoS) components. Additionally, the steering vectors $\mathbf{a}_t(\theta_{p}^t)$ and $\mathbf{a}_r(\theta_{p}^r)$ are defined as the array responses at the BS and the user, respectively. Furthermore, $\theta_{p}^t$ and $\theta_{p}^r$ represent the angle of
departure (AoD) at the BS and the angle of arrival (AoA) at the user, respectively. For a ULA, $\mathbf{a}_t(\theta_{p^t})\in \mathbb{C}^{N_t\times 1}$ and $\mathbf{a}_r(\theta_{p^r})\in \mathbb{C}^{N_r\times 1}$ can be expressed as
\begin{align}
\label{h2}
\mathbf{a}_t(\theta_{p}^t) = \frac{1}{\sqrt{N_t}} [1, e^{-j2\pi \frac{d}{\lambda} \sin{\theta_{p}^t}},\cdot \cdot \cdot, e^{-j2\pi \frac{d}{\lambda} (N_t - 1) \sin{\theta_{p}^t}}]^T,
\end{align}

\begin{align}
\label{h3}
\mathbf{a}_r(\theta_{p}^r) = \frac{1}{\sqrt{N_r}} [1, e^{-j2\pi \frac{d}{\lambda} \sin{\theta_{p}^r}},\cdot \cdot \cdot, e^{-j2\pi \frac{d}{\lambda} (N_r - 1) \sin{\theta_{p}^r}}]^T,
\end{align}
Here, $d$ is supposed as the antenna spacing, while the wavelength of the carrier frequency is defined by $\lambda$. As reported in \cite{gao2015}, $\mathbf H$ has low-rank characteristic since the limited scattering feature in the mmWave massive MIMO channel, indicating that near-optimal throughput is achieved by leveraging limited amounts of RF chains.

Then, we assume a high-dimensional analog precoder and a low-dimensional digital precoder as $\mathbf{D}_A\in \mathbb{C}^{N_t\times N_t^{RF}}$ and $\mathbf{D}_D\in \mathbb{C}^{N_t^{RF}\times N_s}$, respectively, and a hybrid decoder is denoted as $\mathbf D = \mathbf{D}_A \mathbf{D}_D\in \mathbb{C}^{N_t\times N_s}$. Hence, the transmitted signal $\mathbf x$ is given as
\begin{align}
\label{h4}
\mathbf x = \mathbf D \mathbf s = \mathbf{D}_A \mathbf{D}_D \mathbf s,
\end{align}
where $\mathbf s\in \mathbb{C}^{N_s\times 1}$ is the source signal with normalized power $\mathbf{E}[\mathbf s \mathbf{s}^H] = \mathbf{I}_{N_s}$, and we assume that $\tu{tr}\{\mathbf D \mathbf{D}^H\} \leq N_s$ to satisfy the constraint of transmit power \cite{x2018}. Subsequently, the received signal vector is defined as
\begin{align}
\label{h5}
\mathbf y & = \mathbf{B}^H \mathbf{H} \mathbf{x} + \mathbf{B}^H \mathbf{n}
\notag\\ & = (\mathbf{B}_D^H \mathbf{B}_A^H) \mathbf{H} \mathbf{D}_A \mathbf{D}_D \mathbf{s} + \mathbf{B}_D^H \mathbf{B}_A^H \mathbf{n},
\end{align}
Here, $\mathbf{n}\sim \mathcal{CN}(0, \sigma^2\mathbf{I}_{N_s})$ denotes the additive white Gaussian noise (AWGN). Additionally, $\mathbf{B}^H = \mathbf{B}_D^H \mathbf{B}_A^H$ is a hybrid combiner, in which $\mathbf{B}_D\in \mathbb{C}^{N_r^{RF}\times N_s}$ and $\mathbf{B}_A\in \mathbb{C}^{N_r\times N_r^{RF}}$ are defined as the digital combiner and the analog combiner, respectively. Note that the analog precoder/combiner is always installed by analog phase shifters, and all elements of $\mathbf{D}_A$ and $\mathbf{B}_A$ are supposed to meet the requirement as
\begin{align}
\label{h6}
|\{\mathbf{D}_A\}_{i,j}| = \frac{1}{\sqrt{N_t}}, |\{\mathbf{B}_A\}_{i,j}| = \frac{1}{\sqrt{N_r}},
\end{align}

In mmWave massive MIMO system, fully utilizing the sparsity of the mmWave channel can greatly enhance the performance of the hybrid precoding \cite{gao2015}, and thus, we employ a state-of-the-art DNN to construct a novel precoding framework.

\section{Proposed Deep-Learning-Based Hybrid Precoding Scheme}

This part provides a model in which deep learning can be adopted in the mmWave massive MIMO for achieving end-to-end highly efficient hybrid precoding. The splendid learning ability of deep learning enables the spatial features to be exploited of the mmWave massive MIMO system and regard the entire system as a black box to capture useful features for hybrid precoding. We develop the proposed DNN framework and describe how the nonlinear operation can be mapped to the hybrid precoder, and then we provide a novel training policy for facilitating the performance of the DNN.

\subsection{Proposed Deep Neural Network Architecture}

Recently, with the aid of deep learning, considerable process has been achieved in a wide range of areas, including natural language processing (NLP), computer vision (CV), automated driving, and so on. Additionally, deep-learning-based methods can be performed by massively concurrent architectures with distributed memory architectures, such as graphics processing units (GPUs), which have been highlighted for their energy efficiency and impressive computational throughput, arousing great interest in industrial communities.

\begin{figure}
  \centering
  \includegraphics[width=87mm]{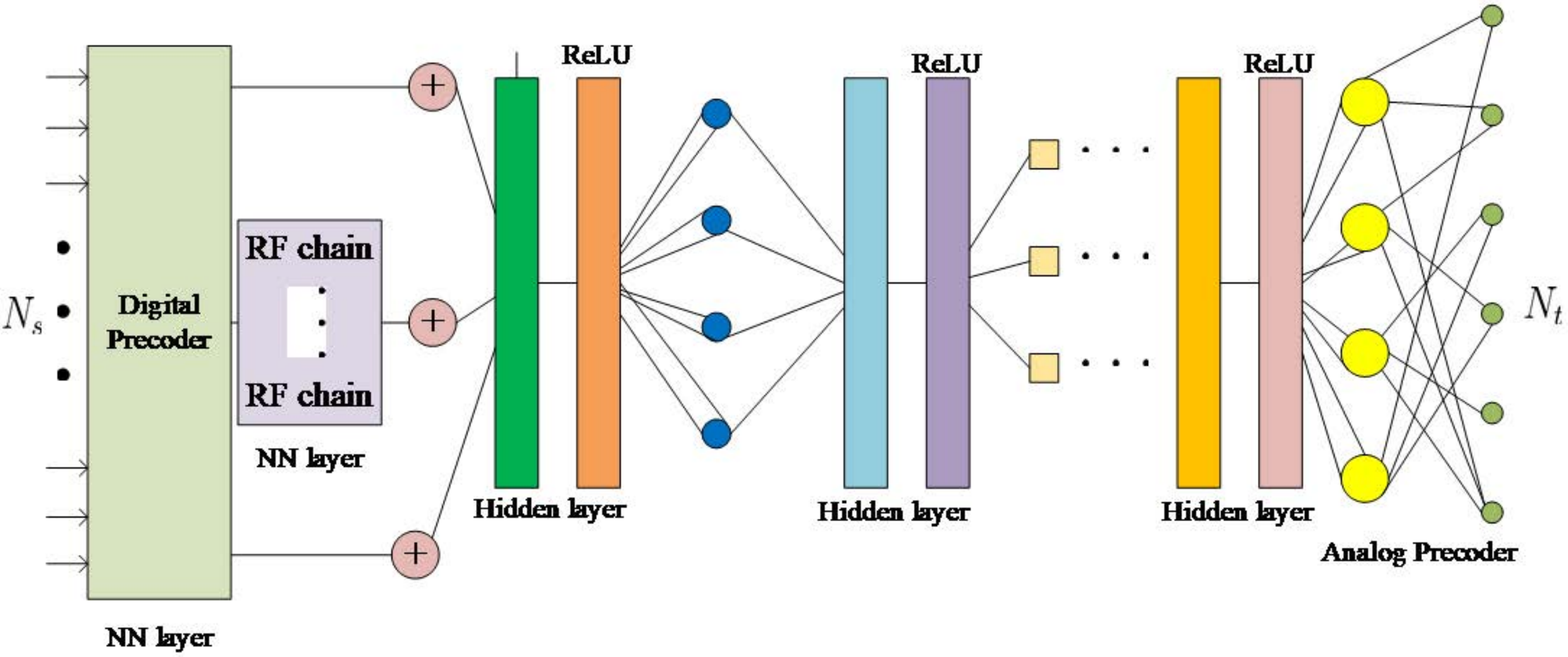}\\
  \caption{DNN architecture in the proposed scheme. }
  \label{fig_auto}
\end{figure}

Deep neural networks (DNNs), the most universal structure of deep learning frameworks, and it can be considered as a multiple layers perceptron (MLP). Specifically, in contrast to an conventional artificial neural network (ANN), many hidden layers are present in a DNN to enhance its learning and mapping abilities. In a DNN, many units are deployed in each hidden layer, and the output can be generated based on the output of these units with the aids of activation functions. In most cases, the rectified linear unit (ReLU) function and the Sigmoid function are used in the nonlinear operation. Assuming $a$ as the argument, they are defined as $\tu{ReLU}(a) = \mathrm{max}(0,a)$ and $\tu{Sigmoid}(a) = \frac{1}{1 + e^{-a}}$, respectively. The output of the network is denoted as $o$ and that $\mathbf{v}$ denotes the input data, the mapping operation can be presented as
\begin{align}
\label{h7}
\mathbf{z} = f(\mathbf{v},w) = f^{(n-1)}(f^{(n-2)}(\cdot\cdot\cdot f^1(\mathbf{v}))),
\end{align}
where $n$ and $w$ represent the number of layers in the neural network and the weights of the neural network, respectively.

To realize hybrid precoding, we construct a DNN framework, as exhibited in Fig. \ref{fig_auto}. Here, in the input layer, the length of each training sequence is determined by its dimension, which is a fully connected (FC) layer with 128 units for capturing features of the input data. The next two hidden layers for processing encoding operation are also FC layers comprising 400 units and 256 units, respectively. Following by this, to disturb the signals with the AWGN or other distortion, a noise layer which is consisted of 200 units for mixing distortion. Subsequently, for the sake of achieving decoding, we design the remaining two hidden layers with 128 units and 64 units, respectively. In addition, the output layer is deployed to generate expected output signals of the network. Moreover, note that ReLU function is introduced as the activation function of the input layer and the hidden layers. However, for enforcing the power constraint in the output layer, a special activation is designed as
\begin{align}
\label{h8}
f(\mathbf{s}) = \min (\max (\mathbf{s}, 0), N_s),
\end{align}

\subsection{Learning Policy}

For simplifying the mapping relation of the hybrid coding, we adopt the GMD method to decompose the complex mmWave massive MIMO channel matrix, and $\mathbf H$ is reformulated by
\begin{align}
\label{h9}
\mathbf y & = \mathbf{W} \mathbf{Q} \mathbf{R}^H
\notag\\ & = [\mathbf{W}_1, \mathbf{W}_2]
\left[ \begin{array}{cc}
\mathbf{Q}_1 & \mathbf{\ast} \\
\mathbf{0} & \mathbf{Q}_2
\end{array}
\right]
\left[ \begin{array}{c}
\mathbf{R}_1^H \\
\mathbf{R}_2^H
\end{array}
\right],
\end{align}
Here, $\mathbf{W}_1\in \mathbb{C}^{N_r\times N_s}$ is a semi-unitary matrix, which is regarded as the combiner. Additionally, $\mathbf{R}_1\in \mathbb{C}^{N_t\times N_s}$ is defined as the precoder, and it is also a semi-unitary matrix. Furthermore, $\mathbf{Q}_1\in \mathbb{C}^{N_s\times N_s}$ is an upper triangular matrix, while $\mathbf{\ast}$ is an arbitrary matrix that can be neglected. Specifically, the largest $N_s$ singular values are formulated as $q_{i,i} = (\delta_1, \delta_2,\cdot \cdot \cdot, \delta_{N_s})^{\frac{1}{N_s}}\in \bar{\mathbf{q}}, \forall i$, where $q_{i,j}$ represent the elements in matrix $\mathbf{Q}_1$. Afterwards, we obtain the received signal as
\begin{align}
\label{h10}
\mathbf y & = \mathbf{B}^H \mathbf{H} \mathbf{x} + \mathbf{B}^H \mathbf{n}
\notag\\ & = \mathbf{W}_1^H \mathbf{H} \mathbf{R}_1 \mathbf{s} + \mathbf{W}_1^H \mathbf{n}
\notag\\ & = \mathbf{Q}_1 \mathbf{s} + \mathbf{W}_1^H \mathbf{n},
\end{align}

To train the hybrid precoder to realize precoding, the loss function is written as
\begin{align}
\label{h11}
loss & = \|\mathbf{R}_1 - \mathbf{R}_A \mathbf{R}_D\|_F
\notag\\ & = \sqrt{\tu{tr}((\mathbf{R}_1 - \mathbf{R}_A \mathbf{R}_D)(\mathbf{R}_1 - \mathbf{R}_A \mathbf{R}_D)^H)}
\notag\\ & = \sqrt{\sum_{i=1}^{min\{N_t,N_s\}} \delta_i^2 (\mathbf{R}_1 - \mathbf{R}_A \mathbf{R}_D)},
\end{align}
where $\|\cdot\|_F$ denotes the Frobenius norm and $\mathbf{R}_A$ and $\mathbf{R}_D$ represent the GMD-based analog precoder and the GMD-based digital precoder, respectively. Additionally, Eq. (\ref{h11}) should satisfy the constraint $|\{\mathbf{R}_A\}_{i,j}| = \frac{1}{\sqrt{N_t}}\}$ and $\tu{tr}(\mathbf{R}_A \mathbf{R}_D \mathbf{R}_D^H \mathbf{R}_A^H)\leq N_s$. Moreover, $\delta_i (\mathbf{R}_1 - \mathbf{R}_A \mathbf{R}_D)$ denote the singular values of matrix $(\mathbf{R}_1 - \mathbf{R}_A \mathbf{R}_D)$.

Next, we employ the DNN framework to construct an autoencoder, which is given by
\begin{align}
\label{h12}
\mathbf{R}_1 = f(\mathbf{R}_A \mathbf{R}_D;\Omega),
\end{align}
$f(\cdot)$ denotes the mapping relation, for which the detailed training procedure is provided as follows, and $\Omega$ is defined as the dataset of the samples.

We consider the proposed deep-learning-based scheme as a mapping operation, and a training mechanism is formulated for extracting the structural statistics of the mmWave-based model. First, we initialize $\mathbf{R}_A$ and $\mathbf{R}_D$ as empty matrices, and then we generate random data sequences in the DNN. Based on different channel conditions, the DNN is trained with the input data sequences, and $\mathbf{R}_A$ and $\mathbf{R}_D$ can be updated. Synchronously, the physical AOA $\theta_{p}^r$ and AOD $\theta_{p}^t$ can be generated randomly, and we can obtain the bias between $\mathbf{R}_1$ and  $\mathbf{R}_A \mathbf{R}_D$ from the output layer of the DNN based on the input signals in different cases through a large number of iterations. Thus, the training dataset $\Omega$ is acquired, consisting of the structural features of the mmWave massive MIMO model and the input data sequences, as well as the output of the DNN. This is an unsupervised learning training approach. In the next stage, the DNN needs to be tested after being trained thoroughly. For each channel condition, the optimal analog precoder $\mathbf{R}_A$ and digital precoder $\mathbf{R}_D$ can be obtained based on the given input signal vectors without requiring iterations. Then, based on the proposed methhod, the stochastic gradient descent (SGD) algorithm with momentum is employed to process the loss function, which is given by
\begin{align}
\label{h13}
\mathbf{R}_A^{j+1} = \mathbf{R}_A^j + v,
\end{align}

\begin{align}
\label{h14}
\mathbf{R}_D^{j+1} = \mathbf{R}_D^j + v,
\end{align}
Here, $v$ is denoted as the velocity for facilitating the gradient element. Additionally, the iteration number is denoted as $j$, and $\mathbf{R}_A^{0}$ and $\mathbf{R}_D^{0}$ are assumed to be the randomly generated initial solution. Specifically, the update procedure of $v$ can be given by
\begin{align}
\label{h15}
v & = \alpha v - \epsilon g
\notag\\ & = \alpha v - \epsilon \frac{1}{N} \bigtriangledown_{\mathbf{R}_A,\mathbf{R}_D} \sqrt{\sum_{i=1}^{min\{N_t,N_s\}} \delta_i^2 (\mathbf{R}_1 - \mathbf{R}_A \mathbf{R}_D)},
\end{align}
where $\alpha$ denotes the momentum parameter and $\epsilon$ denotes the learning rate. Synchronously, $g$ and $N$ represent the gradient element and the number of samples, respectively. Concretely, the learning framework for super hybrid precoding is described in Algorithm \ref{alg:1}.

In addition, to investigate the precoding performance of the deep learning-based precoding strategy, we introduce the mean square error (MSE) to analyze its performance, which can be given as
\begin{align}
\label{h25}
\tu{MSE} = \mathbb{E} \|\mathbf{R}_1 - \mathbf{R}_A \mathbf{R}_D\|^2,
\end{align}

\subsection{Complexity Analysis}

One of the key advantages of the proposed hybrid precoding mean is that this method lowers the computational complexity. Noted that the matrix multiplication is a $N_s \times N_t^2$ space, the complexity of the deep-learning-based method is only $\mathcal{O}(L^2 N_s N_t^2)$, achieving technical advancement compared with that of previous researches, such as the conventional SVD based method. To verify the low computational complexity of the deep-learning-based scheme intuitively, we define $K$ as the number of users, and we present the computational complexity of the proposed method and that of other typical precoding approaches, which is illustrated as TABLE \ref{tab_1}.

\begin{algorithm}
	\renewcommand{\algorithmicrequire}{\textbf{Input:}}
	\renewcommand{\algorithmicensure}{\textbf{Output:}}
	\caption{DNN based hybrid precoding algorithm in massive MIMO.}
	\label{alg:1}
	\begin{algorithmic}[1]
		\REQUIRE The physical AOA $\theta_{p}^r$ and AOD $\theta_{p}^t$, environment simulator.
		\ENSURE Optimized precoder $\mathbf{R}_1$.
		\STATE Initialization: The amount of iteration is initialed as $j = 0$ and the weight is $w = 0$. Meanwhile, initialize error threshold as $\tau = 10^{-7}$. Furthermore, we set $\mathbf{R}_A = \mathbf 0$ and $\mathbf{R}_D = \mathbf 0$.
        \STATE Produce a series of training sequences. Also, $\theta_{p}^r$ and $\theta_{p}^t$ are generated randomly.
        \STATE Construct the proposed DNN framework.
		\STATE Process the environment simulator to simulate wireless channel with artificial distortion or noise.
		\STATE \textbf{while} $(\tu{error}\ge \tau)$\textbf{:} Train the DNN by processing the SGD with momentum according to Eq. (\ref{h13}), Eq. (\ref{h14}), and Eq. (\ref{h15}).
		\STATE Update $\mathbf{R}_A$ and $\mathbf{R}_D$.
        \STATE Obtain the bias between $\mathbf{R}_1$ and $\mathbf{R}_A \mathbf{R}_D$ from the output layer of the network.
        \STATE \textbf{end while}
		\STATE \textbf{return:} Optimized precoder $\mathbf{R}_1$.
	\end{algorithmic}
\end{algorithm}

\begin{table}
\centering
  \caption{Computational complexity of several precoding schemes of mmWave massive MIMO.}
\begin{tabular}{|c|c|}
  \hline
   Methods&Computational complexity\\
  \hline
   Proposed deep learning based scheme&$\mathcal{O}(L^2 N_s N_t^2)$\\
  \hline
   SIC-based hybrid precoding scheme\cite{gao2016}&$\mathcal{O}(2 N_r N_s (N_t^2 (K + N_s N_r)))$\\
   \hline
   Spatially sparse precoding method \cite{o2014}&$\mathcal{O}(2 N_r^3 (N_r^4 N_t + N_r^2 L^2 + N_r^2 N_t^2 L))$\\
   \hline
\end{tabular}
\label{tab_1}
\end{table}

\begin{figure}
  \centering
  \includegraphics[width=87mm]{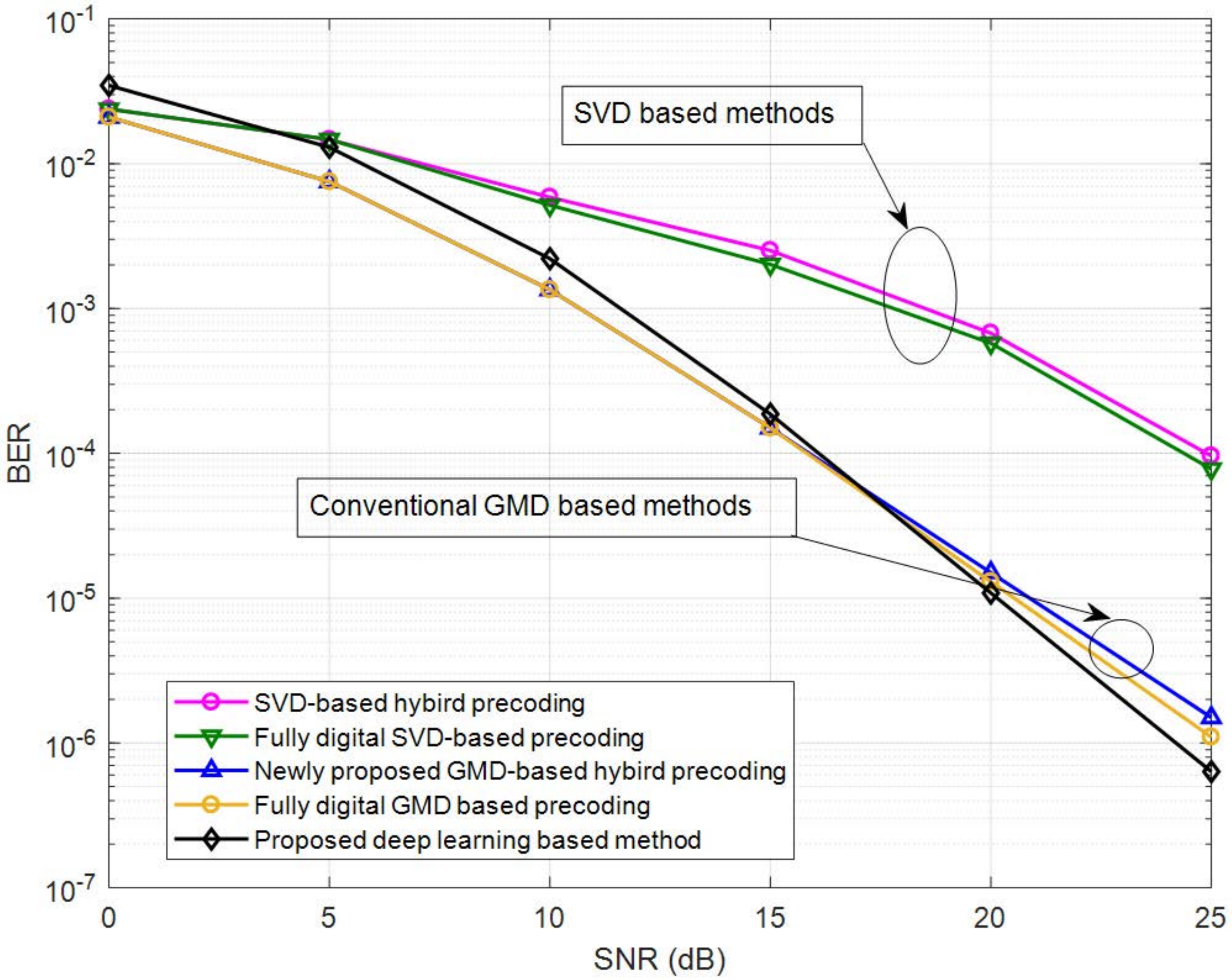}\\
  \caption{BER versus SNR in the case of the proposed DNN-based scheme, SVD-based hybrid precoding scheme \cite{o2014}, fully digital SVD-based precoding method, fully GMD-based precoding method, and new GMD-based precoding scheme \cite{x2018}.}
  \label{fig_compare}
\end{figure}

\begin{figure}
  \centering
  \includegraphics[width=87mm]{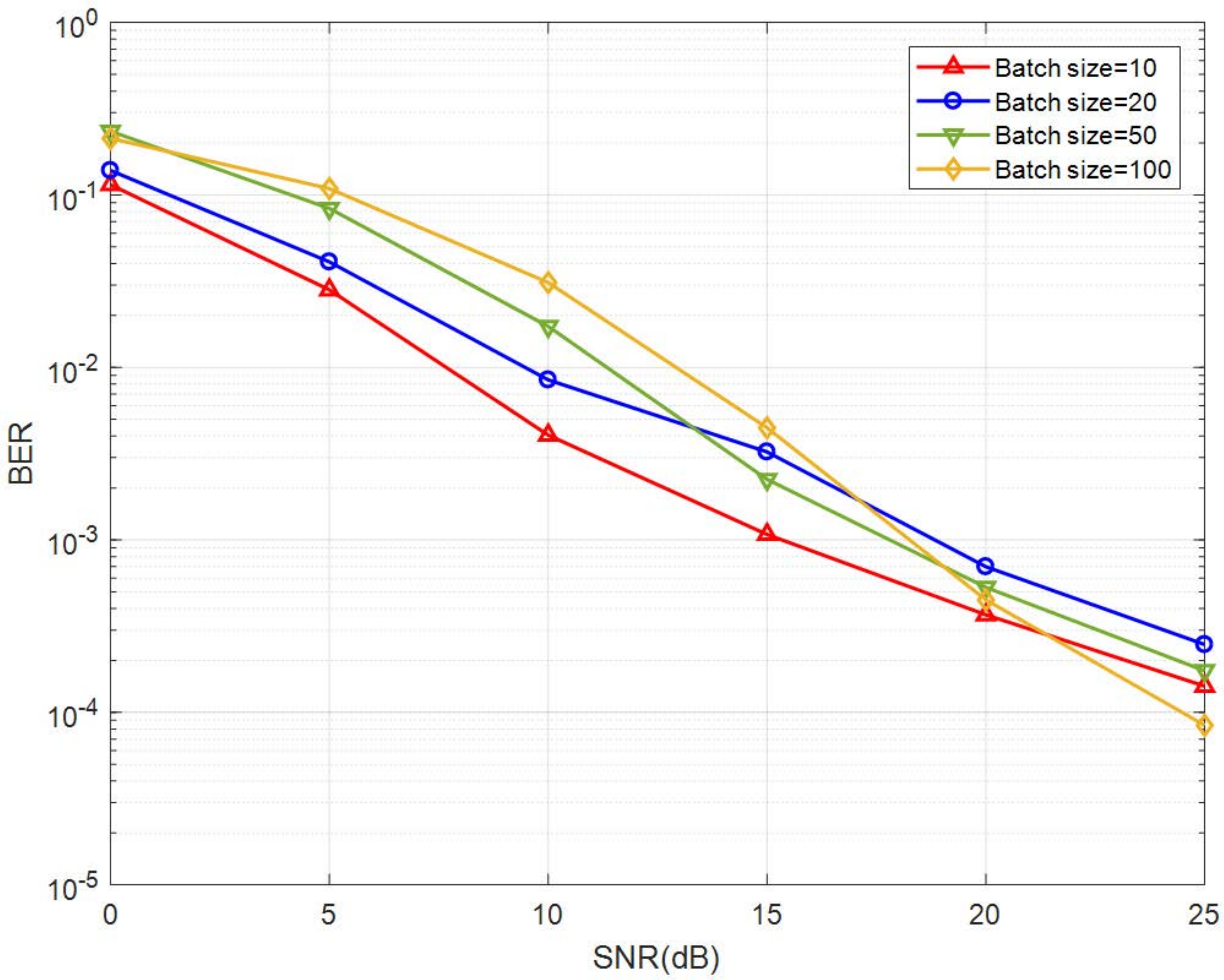}\\
  \caption{BER versus SNR with in the proposed method when the batch size is 10, 20, 50, and 100.}
  \label{fig_batch}
\end{figure}

\section{Numerical Results and Analysis}

In this section, we assess the performance of our proposed DNN-based mmWave massive MIMO scheme with numerical analysis. Here, \emph{Keras} is used to construct and process the DNN framework. Without loss of generality, we generate the channel model according to the models derived in \cite{A2014}. Additionally, $P = 3$, and the carrier frequency is given as 28 GHz. Specifically, the bit error ratio (BER) performance is evaluated with different learning rates and various batch sizes of the training dataset, and its performance is compared with several typical methods. Additionally, the network has been trained for 45000 iterations in the simulation.

To evaluate the superiority of the proposed approach, we investigate the BER performance of the DNN-based scheme compared with those of the SVD-based hybrid precoding scheme \cite{o2014}, fully digital SVD-based precoding method, fully GMD-based precoding method, and new GMD-based precoding scheme \cite{x2018}. As shown in Fig. \ref{fig_compare}, the deep-learning-based method outperforms the conventional schemes. Furthermore, the performance improvement is more apparent between the deep-learning-based strategy and conventional methods, which is attributed to the excellent representation ability of  deep learning.
 Additionally, since the DNN utilizes the structural information and can approach each iteration of the algorithm for hybrid precoding, it is verified that the proposed mmWave massive MIMO strategy is superior to the fully GMD-based digital precoding mean, implying that the existing non-convex optimization in hybrid precoding can be solved with the aid of deep learning.

In Fig. \ref{fig_batch}, the performance of the proposed mmWave massive MIMO scheme is thoroughly investigated when the BER performance is evaluated with various batch sizes. It is observed that the performance of the deep-learning-based strategy degrades with increasing batch size in terms of BER, for the reason that slower convergence may be induced by a larger batch size. However, we further observe that too small batch size will lead to unstable convergence. Therefore, it is noted that we should choose batch size carefully for achieving the optimal performance in the proposed precoding mean.

\begin{figure}
  \centering
  \includegraphics[width=87mm]{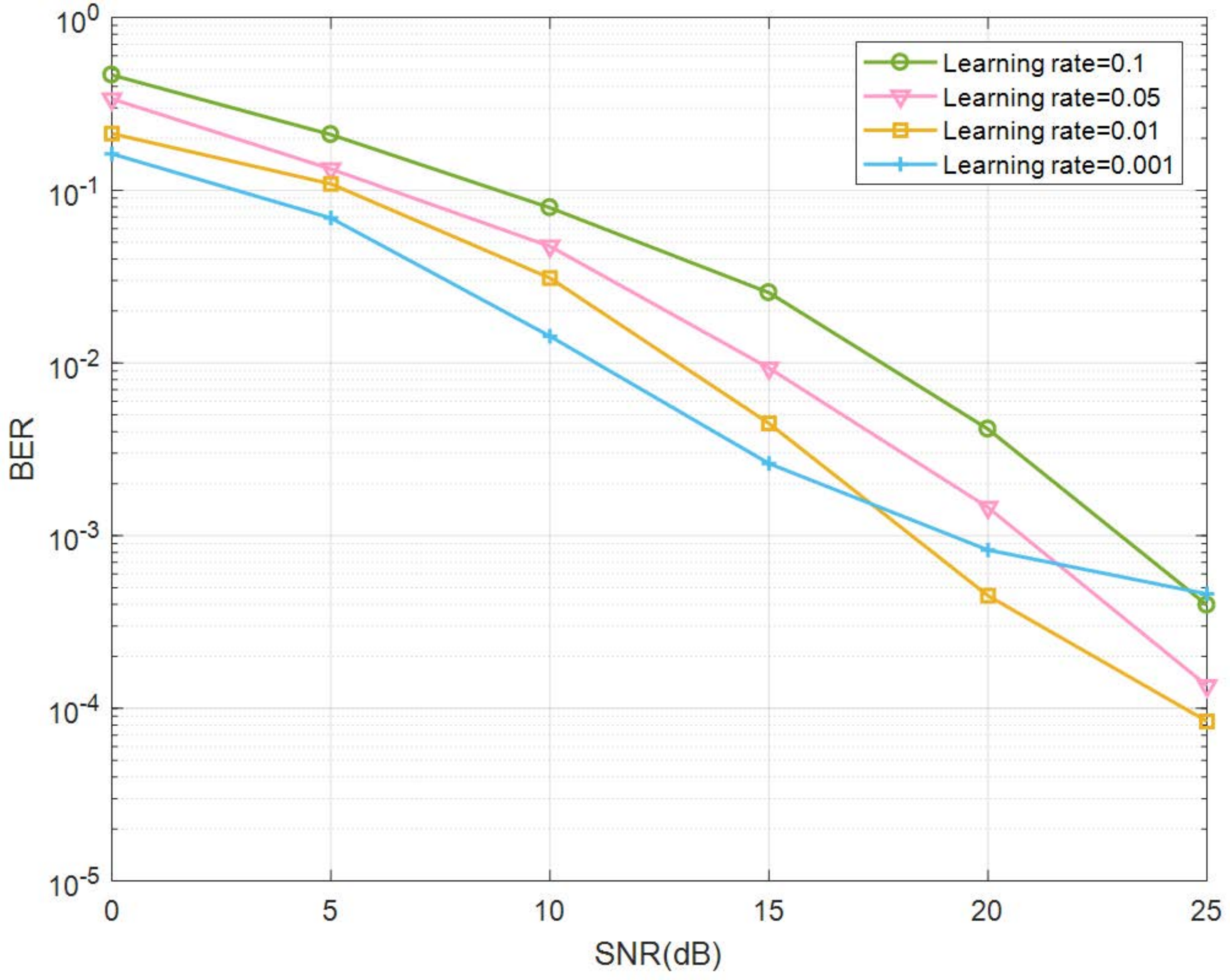}\\
  \caption{BER performance versus SNR in the proposed method with various learning rates.}
  \label{fig_learn}
\end{figure}

\begin{figure}
  \centering
  \includegraphics[width=87mm]{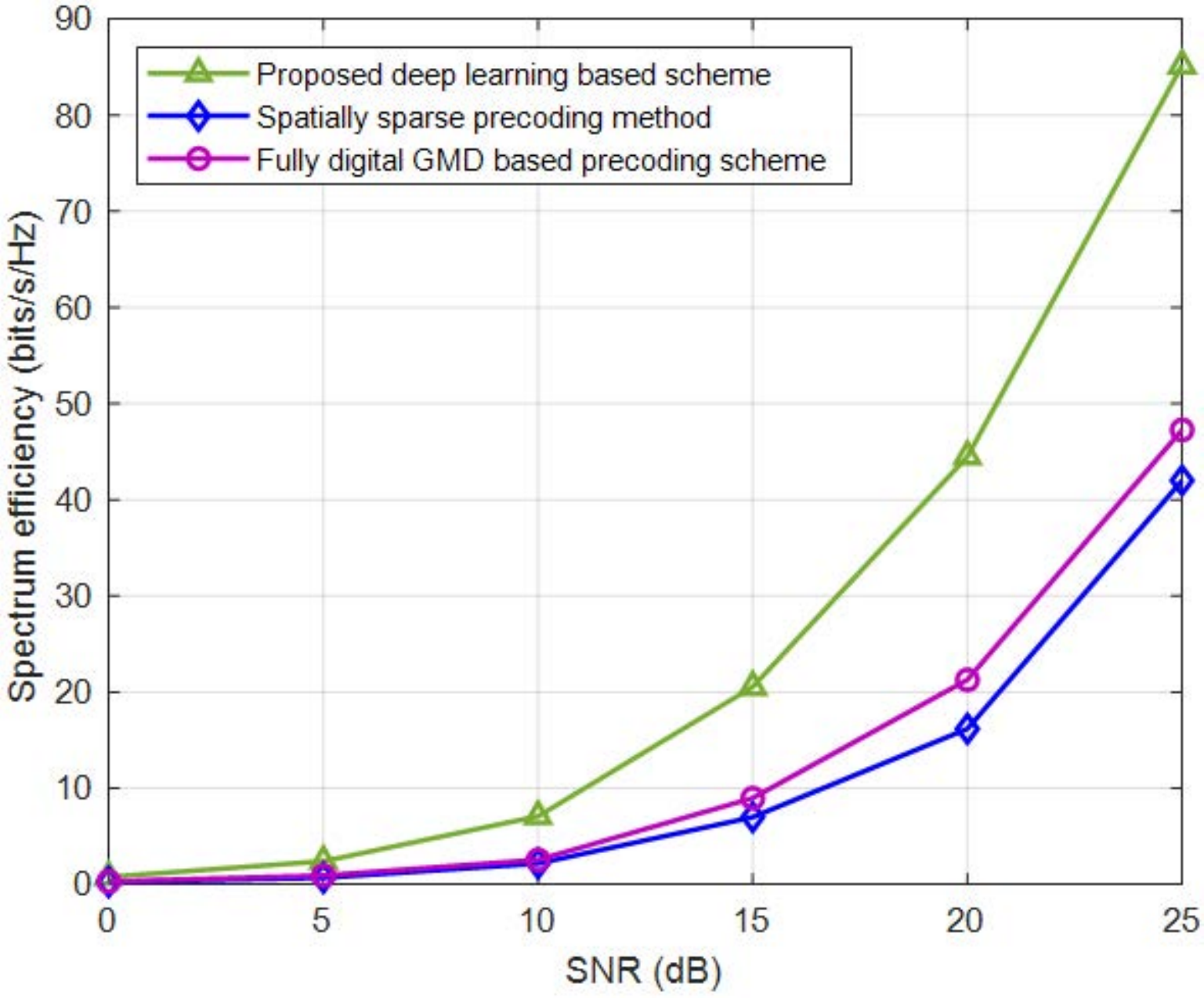}\\
  \caption{Spectrum efficiency versus SNR in the case of the proposed deep learning based hybrid precoding scheme, the spatially sparse precoding method \cite{o2014}, and the fully digital GMD based precoding scheme.}
  \label{fig_sep}
\end{figure}

Fig. \ref{fig_learn} presents the BER versus SNR in the DNN-based mmWave massive MIMO scheme with different learning rates. It can be witnessed in Fig.~\ref{fig_learn} that the performance of hybrid precoding in the DNN-based method is optimized by adopting a lower learning rate, which occurs because a larger learning rate causes a higher validation error. Note, however, that slower convergence behavior will be induced by using a lower learning rate, though it does enhance the system performance. Hence, in order to realize better performance, how to select the best learning rate is an open issue in the proposed framework for hybrid precoding.

In Fig. \ref{fig_sep}, we show the spectrum efficiency performance against the SNR of the DNN-based hybrid precoding scheme, the spatially sparse precoding method \cite{o2014}, and the fully digital GMD based precoding scheme. As shown in Fig.~\ref{fig_sep}, we observe that the spectrum efficiency is improving as the SNR increases in all the schemes. Also, it can be seen from Fig.~\ref{fig_sep} that the proposed hybrid precoding scheme outperforms other strategies, which achieves better hybrid precoding performance dedicated by the excellent mapping and learning capacities of the deep learning. Furthermore, when the SNR increases, the performance gap of the deep learning based scheme and that of other approaches is becoming larger. This superior performance further demonstrates the effectiveness of the proposed hybrid precoding scheme.

\begin{figure}
  \centering
  \includegraphics[width=87mm]{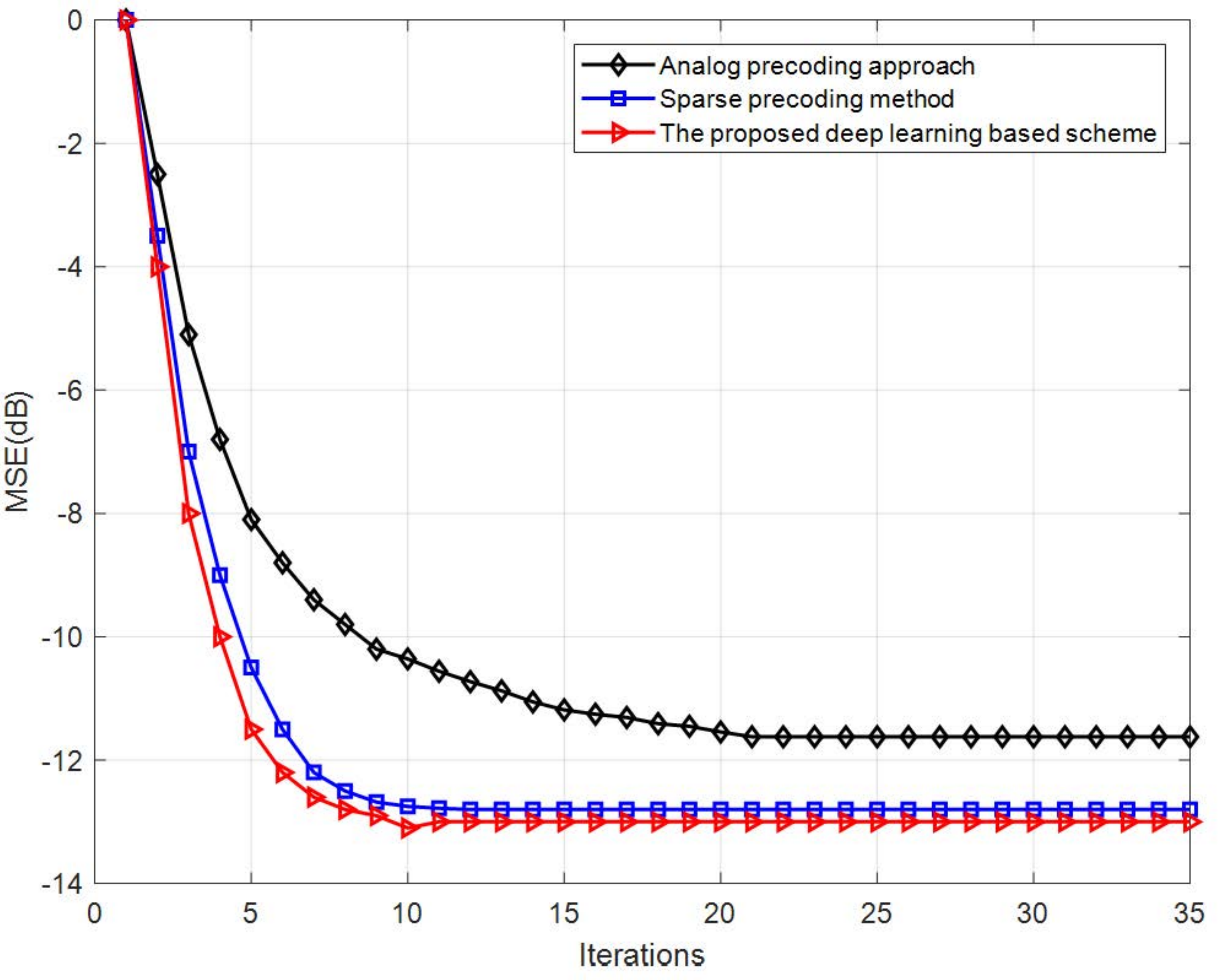}\\
  \caption{MSE versus iterations in the case of the deep learning-based method, the analog precoding approach, and the sparse precoding method.}
  \label{fig_dB}
\end{figure}

Finally, for the purpose of investigating the robustness and stability of the proposed hybrid precoding approach, we explore the relationship between the MSE and the iterations of the deep learning-based strategy compared with the analog precoding scheme, and the sparse precoding method. Here, the learning rate is set as 0.001. As can be seen from Fig. \ref{fig_dB}, we observe that the MSE performance is stirred with the increasing iterations, which is dedicated by the fact that all these algorithms are approaching to conversion with more iterations. Also, it can be seen from Fig.~\ref{fig_dB} that the proposed deep learning based scheme and the sparse precoding method both convert at around 11 iterations, whereas the analog precoding scheme requires about 22 iterations. Furthermore, we can further observe that the MSE performance of the proposed deep learning based scheme is superior than that of other means. Hence, it comes to a conclusion that the proposed deep learning-based approach realizes superior performance in terms of the hybrid precoding accuracy and conversion compared to other schemes.

\balance

\section{Conclusions}

In this paper, we considered a super approach in enhancing the hybrid precoding performance, including cutting down the computational complexity and leveraging the spatial statistics of the large antenna systems in mmWave massive MIMO scenario. We first provided a detailed deep-learning-based hybrid precoding method. Analytical results were presented to verify the performance of the DNN-based method, revealing that the DNN can facilitate the hybrid precoding because of its great recognition and mapping abilities. Another promising direction is to apply deep learning in the channel feedback issue to alleviate the issues related to codebook size and feedback overhead.



\end{document}